\documentclass[twocolumn,showpacs,preprintnumbers,amsmath,amssymb]{revtex4}
\usepackage{epsfig}
\usepackage{graphicx}
\usepackage{dcolumn}

\newcommand{\beq}{\begin{equation}} \newcommand{\eeq}{\end{equation}}
\newcommand{\beqa}{\begin{eqnarray}}
\newcommand{\eeqa}{\end{eqnarray}} \newcommand{\lam}{\lambda}
 \newcommand{\rh}{\rho}
\newcommand{\ga}{\gamma} 
 
 \newcommand{\si}{\sigma}
 \newcommand{\om}{\omega}   
\newcommand{\ra}{\rangle}

\def\josab#1{{ J. Opt.\ Soc.\ Am.\ B\/} {\bf#1}}

\def\pra#1{{ Phys.\ Rev. A\/} {\bf#1}}
\def\prb#1{{ Phys.\ Rev. B\/}{\bf#1}}

\def\prl#1{{ Phys.\ Rev.\ Lett.} {\bf#1}}
\def\sci#1{{ Science} {\bf#1}}

\begin{document}

\date{24 Dec  2005}

\title{Sudden Death of Entanglement: Classical Noise Effects }

\author{T.\ Yu}
\email{ting@pas.rochester.edu}
\author{J.\ H.\ Eberly}
\email{eberly@pas.rochester.edu} \affiliation{Rochester Theory
Center for Optical
Science and Engineering\\
and the Department of Physics \& Astronomy\\
University of Rochester, Rochester, New York 14627}

\begin{abstract}
When a composite quantum state interacts with its surroundings,
both quantum coherence of individual particles and quantum
entanglement will decay. We have shown that under vacuum noise,
i.e., during spontaneous emission, two-qubit entanglement may
terminate abruptly in a finite time [T. Yu and J. H. Eberly, \prl
{93}, 140404 (2004)], a phenomenon termed entanglement sudden
death (ESD).  An open issue is the behavior of mixed-state
entanglement under the influence of classical noise. In this paper
we investigate entanglement sudden death as it arises from the
influence of classical phase noise on two qubits that are
initially entangled but have no further mutual interaction.
\end{abstract}

\maketitle  



\section{Introduction}

On the occasion of Bruce Shore's recent 70th birthday, we are
pleased to join the Quantum Optics community in a celebration of
his career, and specifically by this paper to extend into the
domain of entanglement his contributions to understanding of the
effects of classical noise on qubits and qutrits. His own and
other works are clearly summarized in Chap. 23 of his well-known
two-volume work \cite{ShoreBooks}.

Recently, quantum entanglement has become one theme of research
connected with proposals for realization of many quantum
information protocols, such as quantum cryptography \cite{qcr},
quantum teleportation and quantum computations \cite{preskill,mc}.
Multi-partite entanglement is a key issue in quantum information
processing (QIP),  and has been under extensive research in the
last few years  \cite{de2,kie, pk,lb,hs}.  So far static
entanglement has been investigated extensively, but dynamic
entanglement  under the influence of environmental noises is what
counts in realistic QIP protocols.  Ideally, one hopes that
entanglement needed for quantum information processing can be
maintained for sufficiently long times to permit designed tasks to
be fulfilled. However, as far as we know, noisy evolution of
entanglement of a quantum system  is largely unexplored \cite{ye,
dio, hal, ye-04, buch-04, ye-05, buch-05}.

In a previous publication \cite{ye-04}, contrary to intuition
based on experience about qubit decoherence, we showed that
entanglement may decrease abruptly and non-smoothly to zero in a
finite time due to the influence of quantum noise, specifically
from vacuum fluctuations. This non-smooth finite-time decay is
called entanglement sudden death (ESD), which is  a new kind of
nonlocal decoherence.   In this paper,  we investigate ESD caused
by interaction with noisy classical environments.  More precisely,
we consider two qubits being affected by pure classical dephasing
noises both collectively and individually. Based on this model, we
show that classical noise can also cause ESD in a class of common
mixed states.

The format of the paper is as follows. In Sec. II, we introduce a
model of open quantum systems interacting with their classical
environments. In Sec. III, Wootters concurrence for mixed states
is briefly reviewed and explicit calculations of an important
class of mixed states are given. In Sec. IV we show that there
exist a family of mixed states that have finite disentanglement
times in the case of classical phase damping noises. In Sec. V we
offer some conclusions.

\section{Classical Noisy environments}
\label{channel}
Consider two entangled qubits under the influence of two different
noise models that have been well studied for single qubits
\cite{previous} and also considered for entangled
  states \cite{ye} previously.  Here we focus on mixed states. We
consider the qubits to be spins, and we subject them to noise in
these two different ways:  (i) we impose a stochastic magnetic
field $B(t)$ on both qubits together, and (ii) we impose
stochastic magnetic fields  $b_A(t)$ and $b_B(t)$ separately on
qubit A and qubit B, respectively.  The noises are assumed to be
statistically independent. For simplicity, spatial inhomogeneity
of the noises is not considered in this paper.  Irreversible
coherence decay and entanglement decay will be unavoidable because
we assume that the qubits are members of an extended QIP network
that is too large to rely on the intricate symmetries that would
be necessary to guarantee protection by decoherence-free
subspaces.

\subsection{One Global Collective Noise}

First consider the two qubits to be affected collectively by a
single stochastic field. The hamiltonian of the qubits plus the
classical noisy field is given by: \beq \label{hamiltonian0}
H(t)=-\frac{1}{2}\mu B(t)(\si_z^A +\si_z^B), \eeq where $\mu$ is
the gyromagnetic ratio, and $\si^{A,B}_z$ are the Pauli matrices
in the standard basis: \beqa\label{basis}
|1\ra_{AB}&=&|++\ra_{AB},\,
|2\ra_{AB}=|+-\ra_{AB},\nonumber\\
|3\ra_{AB}&=&|-+\ra_{AB},\, |4\ra_{AB}=|--\ra_{AB}, \eeqa where
$|\pm\pm\ra_{AB}$ denote the eigenstates of the product Pauli spin
operator $\sigma_z^A\otimes\si_z^B$ with eigenvalues $\pm 1$. For
simplicity, we assume that $B(t)$ satisfies the Markov condition:
\beqa <B(t)> & = & 0,\label{cor0}\\
<B(t)B(t')> & = & \frac{\Gamma}{\mu^2}\delta(t-t'),\label{cor1}
\eeqa where $<...>$ stands for an ensemble average and $\Gamma$
gives the dephasing damping rate due to the collective interaction
with $B(t)$.

The solution for the dynamic evolution under the Hamiltonian
(\ref{hamiltonian0}) can be obtained in several different ways
(master equation, stochastic Schr\"{o}dinger equation, etc.), and
we use the operator-sum (Kraus) representation \cite{ye-04}. The
reduced density matrix for the two qubits together can be obtained
from the statistical density operator $\rh_{st}(t)$ for both
qubits and a classical Gaussian field by taking the ensemble
average over the noise field $B(t)$: \beq
\rh(t)=<\rh_{st}(t)>,\label{solution} \eeq where the statistical
density operator $\rh_{st}(t)$ is given by \beq \label{sden}
\rh_{st}(t)=U(t)\rh(0)U^\dag(t), \eeq with the unitary operator
$U(t)=\exp{[-i\int^t_0 dt'H(t')]}.$ Clearly, the unitary operator
$U(t)$ is dependent on the noise $B(t)$.

In our case the statistical unitary operator $U(t)$ can be
explicitly written as \beq
U(t)=\exp\Big[i\frac{\mu}{2}\int^t_0dt'B(t')(\sigma_z^A+\sigma_z^B)\Big]
\eeq By taking the statistical mean of Eq. (\ref{sden}) over the
noise $B(t)$, it can be shown that the most general solution
(\ref{solution}) can be expressed in a very compact way in terms
of Kraus operators \cite{future}: \beq \label{globalsoln}
\rh(t)={\mathcal E}_D(\rh(0)) = \sum^{3}_{\mu=1}D_\mu^\dag(t)
\rh(0) D_\mu(t), \eeq where the Kraus operators describing the
collective interaction are given by \beq \label{model1}
D_1=\begin{pmatrix}
\ga & 0 & 0 & 0\\
0& 1 &  0 & 0\\
0 & 0& 1& 0\\
0 & 0& 0 & \ga
\end{pmatrix},
\eeq \beq\label{model2} D_2=\begin{pmatrix}
\om_1 & 0 & 0 & 0\\
0& 0 &  0 & 0\\
0 & 0& 0& 0\\
0 & 0& 0 & \om_2
\end{pmatrix},
\eeq \beq\label{model3} D_3=\begin{pmatrix}
0 & 0 & 0 & 0\\
0& 0 &  0 & 0\\
0 & 0& 0& 0\\
0 & 0& 0 & \om_3
\end{pmatrix},
\eeq \beqa \ga&=&e^{-t/2T_2},\,\,
\om_1=\sqrt{1-e^{-t/T_2}},\label{para1}\\
\om_2&=&-\om_1 e^{-t/T_2},\,\, \om_3=\om_1^2\sqrt{1+e^{-t/T_2}}
\eeqa where $T_2=1/\Gamma$ is the phase relaxation time due to the
collective interaction with $B(t)$.

\subsection{Two Local Noises}

For the local dephasing model in which two qubits interact with
their own environments represented by two independent classical
noises, the Hamiltonian of the two-qubit system plus the classical
noises is given by: \beq \label{hamiltonian}
H(t)=-\frac{1}{2}\mu(b_A(t)\si_z^A +b_B(t)\si_z^B), \eeq where the
noises $b_A(t)$ and $b_B(t)$ are assumed to be statistically
independent and satisfy: \beqa
<b_i(t)>&=&0, \\
< b_i(t)b_i(t')>&=&\frac{\Gamma_i}{\mu^2}, \delta(t-t'),\,\,
i=A,B.\label{cor2a} \eeqa where  $\Gamma_{i}\,\,\,(i=A,B)$ are the
phase damping rates of qubits A and B due to the coupling to the
stochastic magnetic fields $b_1(t), b_2(t)$, respectively.

Similar to the collective noise case, the general solution of
density matrix $\rho(t)$ of the two qubits can be expressed in
terms of four Kraus operators: \beq \label{kraus2}
\rh(t)={\mathcal E}_{AB} (\rh(0))=\sum^{2}_{\mu, \nu=1}
E^\dag_\mu(t) F^\dag_\nu(t)\rh(0)F_{\nu}(t) E_\mu(t), \eeq where
the Kraus operators describing the interaction with the local
environments are given by \beqa \label{k1} E_1
&=&\left(\begin{array}{clcr}
1 && 0\\
0 && \gamma_A\\
\end{array}
\right)\otimes I,\,\,\, E_2=\left(
\begin{array}{clcr}
0 \,\,&  \,\,\,\, 0 \\
0 \,\,& \,\,\,\om_A\\
\end{array}
\right)\otimes I,\\
F_1&=&I\otimes \left(
\begin{array}{clcr}
1 \,\,&\,\,\, 0\\
0 \,\,&\,\,\, \ga_B\\
\end{array}
    \right),\,\,\, F_2=I\otimes \left(
\begin{array}{clcr}
0 \,\,&  \,\,\, 0 \\
0 \,\,& \,\,\,\om_B\\
\end{array} \right).\label{k5}
\eeqa The parameters appearing in (\ref{k1})--(\ref{k5}) are given
by \beqa \
\ga_A&=&{e}^{-{t}/{2T^A_2}},\,\,\,\ga_B={e}^{-{t}/{2T^B_2}},\\
\om_A&=&\sqrt{1-e^{-{t}/{T^A_2}}},\,\,\,\,
\om_B=\sqrt{1-e^{-{t}/{T^B_2}}}, \eeqa where  $T^A_2=1/\Gamma_A$
and $T^B_2=1/\Gamma_B$ are the phase relaxation times for qubit A
and qubit B due to the interaction with their own environments
$b_A(t), b_B(t)$, respectively.


\section{Measuring Entanglement}
\label{entanle}
To describe the dynamic evolution of quantum entanglement we need
a concrete measure of the degree of entanglement contained in a
quantum state. For any two-qubit case Wootters concurrence
\cite{woo} is particularly convenient. Any reliable measure of
entanglement will yield the same conclusions.  The concurrence
varies from $C=0$ for a separable state to $C=1$ for a maximally
entangled state. For two qubits, the concurrence may be calculated
explicitly from the density matrix $\rho$ for qubits A and B: \beq
C(\rh)=\max\left(0,\sqrt{\lam_1}-\sqrt{\lam_2}-\sqrt{\lam_3}-\sqrt{\lam_4}\,\,\right),
\eeq where the quantities $\lam_i$ are the eigenvalues in
decreasing order of the matrix \beq \zeta=\rho(\sigma^A_y\otimes
\sigma^B_y)\rho^*(\sigma^A_y\otimes \sigma^B_y),
\label{concurrence} \eeq where $\rh^*$ denotes the complex
conjugation of $\rh$ in the standard basis (\ref{basis}) and
$\si_y$ is the Pauli matrix expressed in the same basis as: \beq
\si^{A,B}_y= \begin{pmatrix}
0 \,\,& \,\,\,-i \\
i \,\,& \,\,\,\, 0
\end{pmatrix}.
\eeq

In the following we will examine the evolution of entanglement
under noise-induced relaxation of a class of bipartite density
matrices having the ``standard" form: \beq \label{mixedrho}
\rho^{AB}=\begin{pmatrix}
a & 0 & 0 & w\\
0 &  b &  z & 0\\
0 &  z^* &  c & 0\\
w^* & 0 &  0 & d
\end{pmatrix}.
\eeq where $a+b+c+d = 1$. This class of mixed state arises
naturally in a wide variety of physical situations (see
\cite{pra}). Particularly, it includes pure Bell states as well as
the well-known Werner mixed state \cite{wer} as special cases.

It may be surprising that this class is invariant under dephasing
evolution. Of course the diagonal elements retain their initial
values, but all other elements only pick up a time-dependent
factor multiplying the initial value, so all the initially zero
elements remain zero. A direct calculation shows that the
concurrence of any mixed state of this type is thus given by \beq
\label{fullsoln} C(\rho^{AB})=2\max\{0, |w(t)|-\sqrt{bc},
|z(t)|-\sqrt{ad}\}. \eeq In order for the entanglement to be zero,
both of the following inequalities must be satisfied: \beq
|w(t)|-\sqrt{bc}\le 0, \quad  |z(t)|-\sqrt{ad} \le 0. \eeq In the
following sections we show that they can both be satisfied for
finite times $t$, i.e., that classical noise on entangled spins is
a case where sudden death can occur.

\section{Entanglement sudden death under classical noise}
\label{deco}

\subsection{ Global dephasing noise}
For this case we return to (\ref{globalsoln}) and use the three
Kraus operators to obtain: \beq \label{collective} \rh(t) =
\sum_{\mu=1}^3 D^\dag_\mu(t)\rh(0)D_\mu(t), \eeq where $D_\mu$ are
given in (\ref{model1}),\,\,(\ref{model2}) and (\ref{model3}).

For the initial mixed states (\ref{mixedrho}), the explicit
solution of (\ref{collective}) in the standard basis (\ref{basis})
can be then expressed as \beqa \label{colledeph}
\rho^{AB}(t) & = & {\mathcal E}_D(\rho^{AB}(0))\nonumber\\
& = &\begin{pmatrix}
a & 0 &  0 & \ga^4 w\\
0 & b &  z& 0\\
0& z^* & c & 0\\
\ga^4 w^*&  0 & 0 & d
\end{pmatrix},
\label{initialm} \eeqa where $\gamma=e^{-t/2T_2}$ is defined in
({\ref{para1}).

\begin{figure}[!t]
\epsfig{file=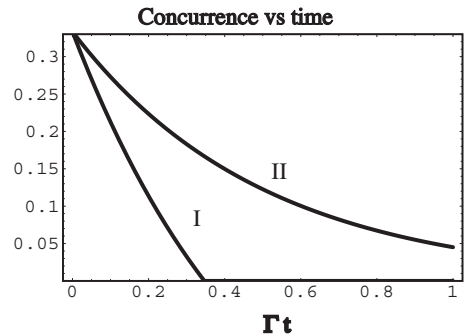, width=6.0 cm}
\caption{{\footnotesize \label{fig1}  In the presence of global
dephasing noise there is long-lived concurrence of mixed entangled
states for $b=0$ or $c=0$.   Otherwise, entanglement sudden death
is unavoidable. The graph shows the sudden death process  (I) with
initial values $z=0, w=1/3, a=d=1/3, b= c=1/6$ and the exponential
decay in (II) with $z=0, w=1/6, a=c=d=1/3, b=0$. Both start from
$C=1/3.$ }}\end{figure}

 From  (\ref{colledeph}), we see that the collective noise only
affects the off-diagonal elements $\rh_{14}$ and $ \rh_{41}$ and
leaves the  off-diagonal elements $z,z^*$ intact. As we noted
previously in discussing pure state decoherence \cite{ye}, the
collective global field allows certain phase combinations to
cancel out, generating what is effectively a decoherence-free
subspace spanned by $|+-\rangle, |-+\rangle$. For our present
mixed-state purpose, we'll avoid this coincidence by assuming that
$z(0)=0$, so the mixed state is not protected in this way.  The
concurrence of the mixed state at $t$ can be easily computed as:
\beq C(\rho^{AB}(t))=2\max\{0, |w(t)|-\sqrt{bc}\} \eeq Therefore,
the state $\rh^{AB}(t)$  (\ref{initialm}) is separable if and only
if  $|w(t)|-\sqrt{bc}\le 0$.  From this we see that there is a
critical time $t_c$ for the end of entanglement, such that \beq
\label{criticaltime}
t_c=\frac{1}{2\Gamma}\ln\frac{|w|}{\sqrt{bc}}, \eeq at which
$C(\rho^{AB}(t_c) = 0$. From  (\ref{criticaltime}), we see that
for the entangled density matrix with $b\neq 0$ and $c\neq 0$, the
entanglement sudden death will occur at $t_c$.  These features are
illustrated in Fig. 1, where the sudden death time is $\Gamma
t_c=\frac{1}{2}\ln 2$.

\subsection{Two-qubit local dephasing noises}

In the case of local dephasing the independence of the two local
noises prevents appearance of a decoherence-free subspace, and the
general solution is given by eqn. (\ref{kraus2}), which here takes
form: \beqa\label{ch} &&\rh(t)=\begin{pmatrix}
\rho_{11} &\ga_B\rho_{12}&  \ga_A\rho_{13} & \ga_A\ga_B\rho_{14}\\
\ga_B\rho_{21} &\rho_{22}&  \ga_A\ga_B\rho_{23}& \ga_A\rho_{24}\\
\ga_A\rho_{31}&\ga_A\ga_B\rho_{32}&\rho_{33}& \ga_B\rho_{34}\\
\ga_A\ga_B\rho_{41}&\ga_A\rho_{42}&\ga_B\rho_{43}& \rho_{44}
\end{pmatrix}.
\eeqa

The general solution (\ref{fullsoln}) easily shows that under
independent phase noises one will always find entanglement sudden
death for the initial mixed density matrix (\ref{mixedrho}).

An interesting sub-category is worth attention. This is the case
when one of the two qubits experiences very weak noise or no
noise. One such limiting case is, for example, $\Gamma_B = 0$. The
one-qubit dephasing model ${\mathcal E}_A$ can be described by the
Kraus operators $E_1, E_2$ and one can write the explicit solution
for any initial state $\rh(0)$, but the result is easy to
anticipate.We find the result (\ref{ch}) again, but have to put
$\Gamma_B = 0$, and otherwise the critical time is the same.
Naturally it is a longer time, but still finite. Again we find
ESD.

As a most striking example,  we have shown that the effect of the
dephasing noises on entanglement  and quantum coherence of a
single qubit is indeed very different. Particularly, we have shown
here that for some mixed states,  entanglement may experience a
sudden death process even if the local coherence of one
participating particle can be well  preserved and the other  only
decays to zero asymptotically. This may be one of the best
examples  to show the difference between entanglement decay and
local coherence decay.

\section{Conclusion}
\label{conc}
In this paper for a standard set of initial mixed two-qubit states
we have investigated quantum entanglement decay due to interaction
with classical noises.  We have shown that noisy classical
environments may cause entanglement to vanish completely in a
finite time, while they allow the coherence of either one or both
of the engaged qubits to remain non-zero for an infinitely long
time. A deeper understanding of  entanglement decay processes is
of interest in  quantum computation and in any branch of quantum
information science where preservation of entanglement is
essential for the action of desired operations and devices.

A few closing comments are in order: (1) The degree of
entanglement in this paper is measured by Wootters' concurrence,
which is defined for both pure and mixed states. However, we
emphasize that entanglement sudden death (ESD) is independent of
the choice of measures of entanglement. This can be easily seen as
follows: Entanglement sudden death occurs when a quantum entangled
state becomes separable beginning at a certain time, but
separability of a state is independent of entanglement measure.
Concurrence is a conveniently normalized measure. (2) We note that
because we have found unexpected evolution of standard mixed
states in the presence of white noise sources, it will be very
interesting  to extend these results to the case of colored noises
(e.g., see \cite{hs}). (3) We note that entanglement sudden death
is a generic feature for a still larger class of  mixed states
than we have specified in this paper. (4) It will of great
interest if analogies of entanglement sudden death can be
identified in multi-partite systems.


\section*{Acknowledgments}

We acknowledge financial support for our research from ARO Grant
W911NF-05-1-0543.


\begin{thebibliography}{10}
\bibitem{future} The details of derivation of the Kraus representation will
be published elsewhere.
\end{thebibliography}
\end{document}